\documentclass{ws-jai}
\usepackage[flushleft]{threeparttable}
\begin{document}

\catchline{}{}{}{}{} % Publisher's Area please ignore

\markboth{Zakharenko V. et al.}{Digital receivers for low-frequency radio telescopes UTR-2, URAN, GURT}

\title{Digital receivers for low-frequency radio telescopes UTR-2, URAN, GURT
%\footnote{For the title, try not to use more than
%three lines. Typeset the title in 11 pt Times Roman,
%boldface, with the first letter of important words capitalized.}
}

\author{
Zakharenko~V.$^{\S\ast}$, Konovalenko~A.$^\S$, Zarka~P.$^\dag$, Ulyanov~O.$^\S$, Sidorchuk~M.$^\S$, Stepkin~S.$^\S$, Koliadin~V.$^\S$, Kalinichenko~N.$^\S$, Stanislavsky~A.$^\S$, Dorovskyy~V.$^\S$, Shepelev~V.$^\S$, Bubnov~I.$^\S$, Yerin~S.$^\S$, Melnik~V.$^\S$, Koval~A.$^\S$, Shevchuk~N.$^\S$, Vasylieva~I.$^\S$, Mylostna~K.$^\S$, Shevtsova~A.$^\S$, Skoryk~A.$^\S$, Kravtsov~I.$^\S$, Volvach~Y.$^\S$, Plakhov~M.$^\S$, Vasilenko~N.$^\S$, Vasylkivskyi~Y.$^\S$, Vavriv~D.$^\S$, Vinogradov~V.$^\S$, Kozhin~R.$^\S$, Kravtsov~A.$^\S$, Bulakh~E.$^\S$, Kuzin~A.$^\S$, Vasilyev~A.$^\S$, Ryabov~V.$^\ddagger$, Zarka~P.$^\dag$, Reznichenko~A.$^\S$, Bortsov~V.$^\S$, Lisachenko~V.$^\S$, Kvasov~G.$^\S$, Mukha~D.$^\S$, Litvinenko~G.$^\S$, Brazhenko~A.$^\dagger$, Vashchishin~R.$^\dagger$, Pylaev~O.$^\dagger$, Koshovyy~V.$^\alpha$, Lozinsky~A.$^\alpha$, Ivantyshin~O.$^\alpha$, Rucker~H.~O.$^\beta$, Panchenko~M.$^\gamma$, Fischer~G.$^\gamma$, Lecacheux~A.$^\dag$, Denis~L.$^\backepsilon$, Coffre~A.$^\backepsilon$, and Grie\ss meier~J.-M.$^\delta$ }
\address{
$^\S$Institute of Radio Astronomy, National Academy of Sciences of Ukraine, Kharkiv, Ukraine, zakhar@rian.kharkov.ua\\
$^\ddagger$Future University Hakodate, Hakodate, Japan\\
$^\dag$LESIA $\&$ USN, Observatoire de Paris, CNRS, PSL/SU/UPMC/UPD/SPC/UO/OSUC, Meudon,
France\\
$^\dagger$Poltava gravimetrical observatory of Institute of geophysics, National Academy of Sciences of
Ukraine, Poltava, Ukraine \\
$^\alpha$Karpenko Physiko-Mechanical Institute, National Academy of Sciences of Ukraine, Lviv, Ukraine\\
$^\beta$Commission for Astronomy of Austrian Academy of Sciences, Wien, Austria\\
$^\gamma$Space Research Institute, Austrian Academy of Sciences, Graz, Austria\\
$^\backepsilon$USN, Nan\c{c}ay, France \\
$^\delta$LPC2E, Orl\'{e}ans, France
}

\maketitle

\corres{$^\ast$Corresponding author.}

\begin{history}
\received{(to be inserted by publisher)};
\revised{(to be inserted by publisher)};
\accepted{(to be inserted by publisher)};
\end{history}

\begin{abstract}
This paper describes digital radio astronomical receivers used for decameter and meter wavelength observations. Since 1998, digitals receivers performing on-the-fly dynamic spectrum calculations or waveform data recording without data loss have been used at the UTR-2 radio telescope, the URAN VLBI system, and the GURT new generation radio telescope. Here we detail these receivers developed for operation in the strong interference environment that prevails in the decameter wavelength range. Data collected with these receivers allowed us to discover numerous radio astronomical objects and phenomena at low frequencies, a summary of which is also presented.

\end{abstract}


\section{Introduction}
\noindent 
Digital receiving equipment performs a large part of data processing and analysis in astronomy. In modern low-frequency (LF) radio telescopes, digitization and subsequent digital processing can start practically at the output of each dipole. Reconfigurable observation modes with high temporal and spectral resolutions, coupled with broad frequency range and multibeam capabilities provided by digital equipment and processing algorithms, have taken radio astronomy research to a new technological level.

New radio telescopes such as LOFAR (Low Frequency Array, the Netherlands, in the range (10)30-80 and 110-240 MHz) \cite{2013A&A...556A...2V}, the LWA (Long Wavelength Array, USA, 10-88 MHz) \cite{2012JAI_Tailor_LWA}, the MWA (Murchison Widefield Array, Australia, 80-300 MHz), NenuFAR (New Extension in Nan\c{c}ay Upgrading LOFAR, France, 10-85 MHz) \cite{2012Zarka_LSS_NenuFAR}, and the future SKA\footnote{Square Kilometer Array, http://astronomers.skatelescope.org} project all utilize digital systems.

The large field of view and high spatial resolution of LF interferometric telescopes such as LOFAR and SKA permit to study various astrophysical targets by building high quality maps. However, this require real-time processing of huge data volumes.

In addition to imaging, various types of sporadic radio emission are studied: pulsars \cite{Coenen2014,Stovall2015,2016A&A...585A.128K} including giant pulses \cite{Tsai2016}, Jupiter \cite{Girard2016}, the Sun \cite{Morosan2015} and other transient signals \cite{2011A&A...530A..80S,Rowlinson2016}. These results show the vast capabilities offered by digital receivers at standalone radio telescopes \cite{Kocz2015} as well as in multi-telescope observations\footnote{http://www.rri.res.in/}.

LF radio observations (below 100 MHz) are of particular astrophysical interest due to non-thermal emission mechanisms and related physical processes that operate at these frequencies. In this range, despite the relatively narrow absolute frequency band, the upper-to-lower frequency ratio often exceeds several octaves. This helps to measure frequency-dependent emission parameters (spectral density, dispersion delay, scattering time constant, etc.) with high accuracy, possibly leading to important astrophysical conclusions.

Modern digital receivers have such a broad simultaneous band that they can record the entire LF range at once (in the baseband or in upper Nyquist zones: $\Delta f = [n,n+1]f_{\rm{sampling}}/2$ with n = 0, 1, 2, \dots). This requires only filtering and eliminates the need for prior frequency conversion, simplifying thus the structure of the receiving system. 

The number of bits of modern analog-to-digital converters (ADC) has increased from 1-4 to 12-16, resulting in a much higher dynamic range. This is especially important for the frequencies below 30 MHz where radio frequency interference (RFI) from broadcasting stations are many orders of magnitude stronger than the weak cosmic signals.

Multiple implementations of digital receivers for LF radio telescopes have been described previously, for example for LOFAR\footnote{http://www.lofar.org} and the LWA \cite{Soriano2011,Warnick2016}. As each telescope has a unique design adapted to its specific observational goals, it is often inappropriate to borrow digital receiver components developed for other purposes without modification. Thus, it is often necessary to develop a dedicated equipment for particular astrophysical problems or for a specific telescope. Conversely, lessons learned in digital receiver development and operation at a given telescope may be helpful elsewhere. This is the motivation that inspired the present paper, that briefly reviews digital receivers developed for the Ukrainian LF radio arrays UTR-2, URAN, and GURT \cite{Konovalenko2016}, which have already delivered important astrophysical results. Section 2 describes the existing Ukrainian LF radio telescopes and those under construction, with their successive generations of digital receivers. The main lessons are that LF radio astronomy instruments should take into account operation in a harsh RFI environment, a strong dependence of the galactic background brightness temperature on the frequency, and losses in the analog phasing systems of antenna arrays. Section 3 presents the capabilities and parameters of operation of the new generation of receivers installed at the above telescopes. Section 4 provides a selection of astrophysical results obtained with these new digital receivers. The last section summarizes the advances and briefly explores the promising avenues for future developments in LF radio astronomy.

\section{Technical background and requirements for low-frequency digital receivers}
\subsection{Antenna systems}

Since the late 1950s, scientific and instrumental advances of the department of Astronomy of the Institute of Radiophysics and Electronics (that later became the Institute of Radio Astronomy) of NAS (National Academy of Sciences) of Ukraine have been focused in the very low frequency range\footnote{corresponds to HF in the ITU classification}. The LF limit is imposed by the Earth's ionosphere, that reflects out incoming radio waves with frequency below a few MHz (the exact cutoff depends on the season and the local time). 

Antenna systems for LF radio telescopes are usually large phased antenna arrays. In order to phase and add the signals, coaxial cable transmission lines were used until now and are still in use in some cases. Electrical losses in these cables strongly depend on the frequency. They are compensated by specific preamplification systems (PAS).

\begin{figure}%[h]												% Fig_UTR-2
\begin{center}
\includegraphics[width=1\linewidth]{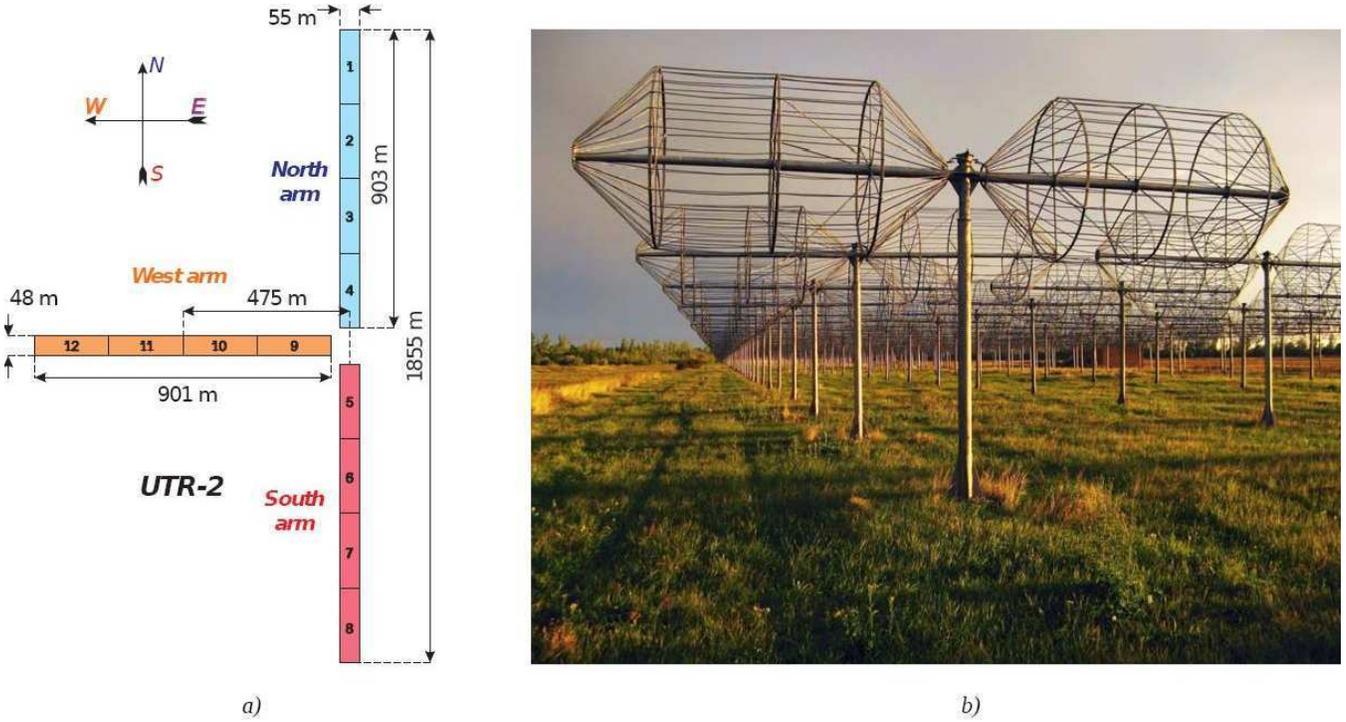}  
\end{center}
\caption{$(a)$ Geometrical configuration of the UTR-2 radio telescope. $(b)$ Picture of antennas of the south arm of UTR-2.}
\label{UTR-2}
\end{figure}

\begin{figure}%[h]												% Fig_URAN-2
\begin{center}
\includegraphics[width=1\linewidth]{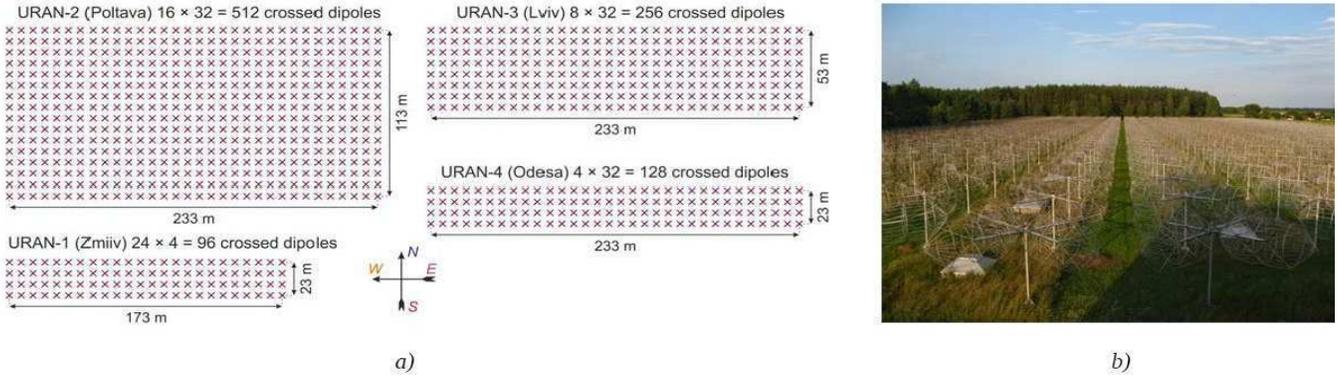}  
\end{center}
\caption{$(a)$ Geometrical configurations of the 4 URAN radio telescopes. $(b)$ Picture of URAN-2 antennas.}
\label{URAN-2}
\end{figure}

\begin{figure}%[h]												% Fig_GURT #3
\begin{center}
\includegraphics[width=1\linewidth]{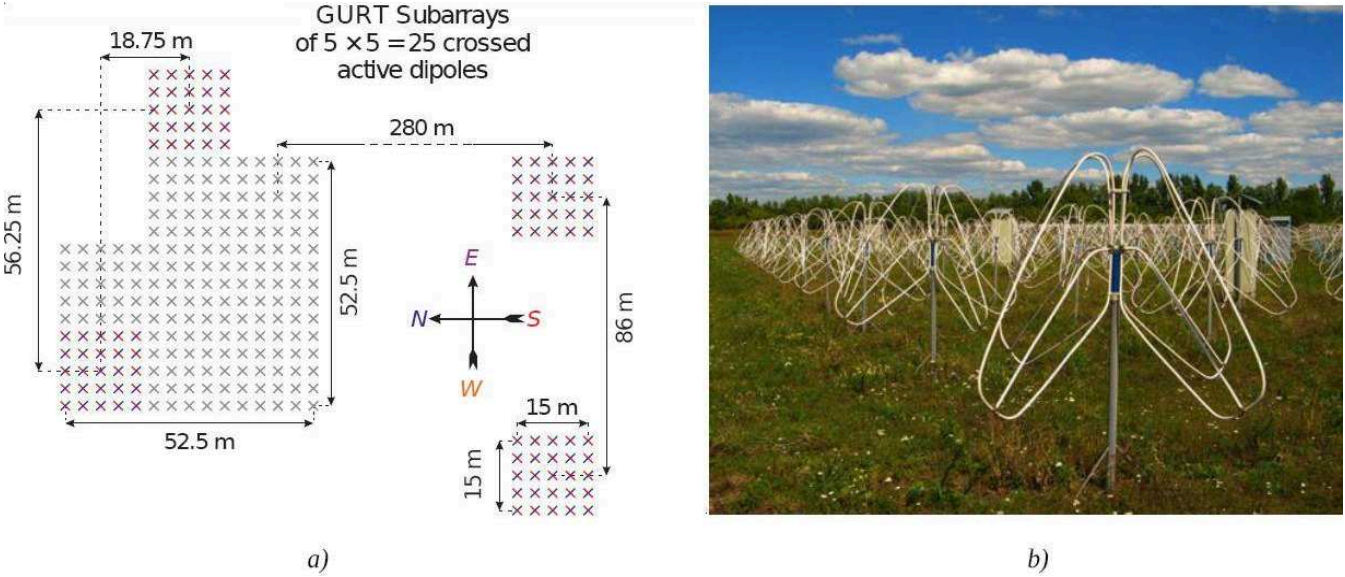}  
\end{center}
\caption{$(a)$ Geometrical configuration of existing GURT subarrays of orthogonal dipoles. Fully commissioned and operational subarrays are displayed in blue and red, whereas grey antennas indicate 
installed subarrays not yet connected to back-ends. $(b)$ Picture of a cluster of GURT subarrays.}
\label{GURT}
\end{figure}

The first Ukrainian prototype of a LF radio telescope was UTR-1 \cite{Bruck1968}, which served as a test bench for technical development and circuit designs. It was followed by the Ukrainian T-shaped radio telescope, version 2 (UTR-2,  \citet{1978Anten..26....3B}, Fig.~\ref{UTR-2}), the URAN-1 to -4 radio interferometer (\citet{Megn1997,Megn2003}, Fig.~\ref{URAN-2}), and recently the first sections of the new generation Giant Ukrainian Radio Telescope (GURT, \citet{Konovalenko2016}, Fig.~\ref{GURT}). The effective area, frequency range, number and polarization of antennas of these telescopes are listed in Table~\ref{tbl1}.

\begin{wstable}[h]												% Table 1
\caption{Parameters of Ukrainian low-frequency radio telescopes.}
\begin{tabular}{@{}cccc@{}} \toprule
Telescope & Frequency range  & Effective area $A_{\rm{eff}}$ at &
Number of antenna elements, \\
& (MHz) & center frequency (m$^2$) & and polarization \\ \colrule
UTR-2\hphantom{00} & \hphantom{0} 8-32 & \hphantom{0}140 000 & 2040, one linear \\
URAN-1\hphantom{00} & \hphantom{0} 10-35 & \hphantom{0}5 500 & 96, two linear \\
URAN-2\hphantom{00} & \hphantom{0} 10-35 & \hphantom{0}28 000 & 512, two linear\hphantom{0} \\
URAN-3\hphantom{00} & \hphantom{0} 10-35 & \hphantom{0}14 000 & 256, two linear\\
URAN-4\hphantom{00} & \hphantom{0} 10-35 & \hphantom{0}7 300 & 128, two linear\\ 
GURT\hphantom{00} & \hphantom{0} 8-80 & \hphantom{0}5 $\times$  350 & 5 $\times$ 25, two linear\\\botrule
\end{tabular}
\label{tbl1}
\end{wstable}

Neglecting confusion, the sensitivity (minimum detectable flux density) of a radio telescope can be calculated using the radiometer equation:
\begin{equation}
\delta S_{\rm{min}} = 2kT/(A_{\rm{eff}} \sqrt{\delta f \delta t}),
\end{equation}
where $k$ is the Boltzmann constant, T the system temperature (dominated by the brightness of the Galactic background), $\delta t$ and $\delta f$ the integration time and bandwidth. For UTR-2 $A_{\rm{eff}}$ = 140 000 m$^2$ and taking $T = 40 000$ K at 20 MHz,  an integration time $\delta t = 1$ h and a bandwidth $\delta f = 3$ MHz, we obtain $\delta S_{\rm{min}} \approx 10$ mJy.

UTR-2 and URAN-1 to URAN-4 have fully analog signal phasing and summation systems, based on time delay lines implemented by the switching of coaxial cable segments. UTR-2 is the decameter wave radio telescope with the largest effective area in the world. The structure of the array and its phasing system are described in detail in \cite{1978Anten..26....3B}. As its name indicates, UTR-2 has a T-shape with three arms: North, West and South (Fig.~\ref{UTR-2}). This configuration combines a large effective area with a good angular resolution (0.5$^{\circ}$). Its analog phasing system provides 2048 beam positions in the North-South direction ($\pm$90$^{\circ}$ from zenith) $\times$ 1024 beam positions in the East-West direction ($\pm$60$^{\circ}$ from the meridian). The high filling factor of UTR-2 ensures low side lobe and diffraction lobe levels. The phasing system is able to form five simultaneous beams in the local meridian plane, shifted by 0.5$^{\circ}$ with respect to each other in the North-South direction. These simultaneous beams provide a limited mapping capability. A 4-stage PAS is used in order to compensate the losses in the phasing system (up to 70 dB at 33 MHz) and ensure maximum signal-to-noise ratio (SNR) and linearity \cite{Abranin1997, Abranin2001}. By splitting the signal into three subbands (8-11, 11-18 and 18-32 MHz) with 12 dB gain drop between consecutive subbands, the PAS delivers a nearly flat spectrum ($\pm$ 6 dB) in the 10-30 MHz range.

\begin{figure}[h]												% Fig_eDay   fig4
	\begin{center}
		\includegraphics[width=0.5\linewidth]{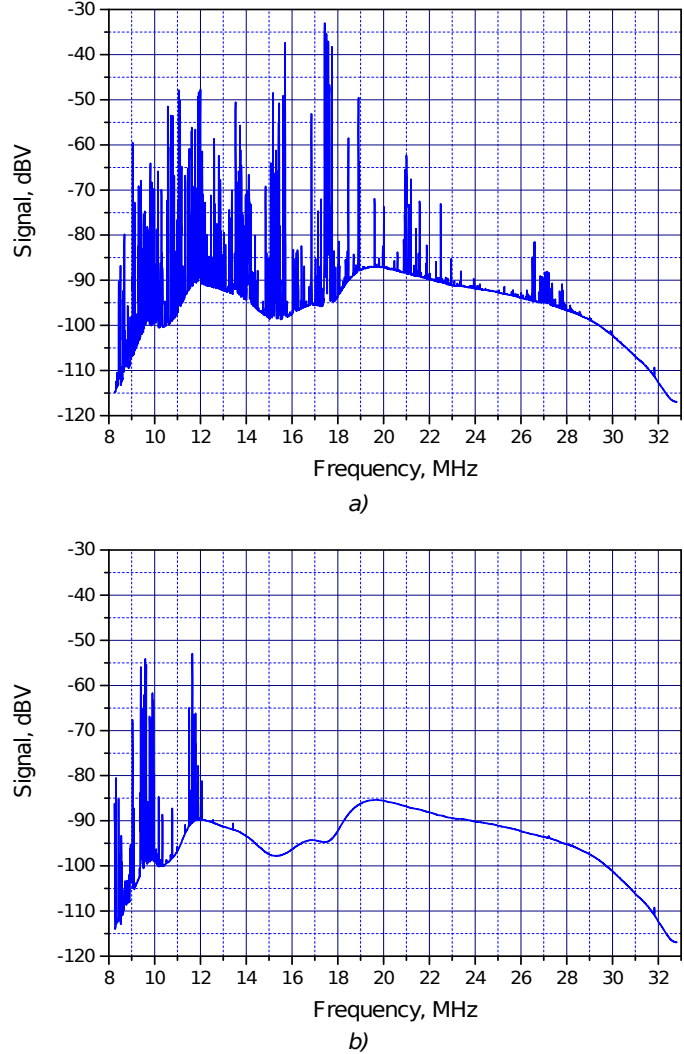}  
	\end{center}
	\caption{Typical spectrum from UTR-2 illustrating the daytime $(a)$ and nighttime $(b)$ RFI conditions. Integration time is 100 s and frequency resolution (and thus integration bandwidth) is 4 kHz.}
	\label{eDay}
\end{figure}

Fig.~\ref{eDay} illustrates typical day and night RFI conditions at UTR-2. The large dynamic range of the electronics results in a very low intermodulation level, much lower than the intrinsic noise of the PAS. This enables long integration times that improve the SNR, which is vital for detecting weak signals from dim astrophysical sources.

While UTR-2 provides measurements in a single linear (EW) polarization, the URAN telescopes have dual-polarized antennas (Fig.~\ref{URAN-2}). Their rectangular shape elongated in the East-West direction and high filling factor (Fig. 2a) result in large effective areas. Their compacity implies lower losses in the phasing system (compared to UTR-2), so that a simpler PAS (with 2 subbands and 2 stages) provides a flat spectrum with high linearity. Together, UTR-2 and URAN telescopes have an effective area $\sim$200Â 000 m$^2$ (0.2 km$^2$).

GURT is a new generation telescope developed in order to observe in a broader frequency range than UTR-2 and URAN, and consequently capable of coordinated simultaneous observations with the American LWA and the European LF arrays: LOFAR's low-band antenna (LBA) array, the Nan\c{c}ay Decameter Array (NDA) \cite{1980Icar...43..399B}, and NenuFAR. GURT, currently under construction, will consist of many ($\sim$100) identical subarrays. It will have a hybrid phasing system, with analog phasing within each subarray (like in UTR-2), and digital phasing between subarrays. The signals from each subarray output will be digitized, and then digitally phased, added and further processed. This approach has several advantages: each subarray is actually a small radio telescope capable of addressing various astrophysical questions (see section 4). Digital subarray phasing is RFI-proof and does not introduce frequency-dependent attenuation to the signal, in contrast to coaxial cables. It also provides flexible beamforming at lower construction and maintenance costs than delay lines. A similar approach is used in e.g. LOFAR's high-band antenna array and NenuFAR.

\subsection{Main challenges and requirements for new digital receivers at low frequencies} % 												Section 2-2

Broadband receivers for LF phased arrays must cope with the large variations of the received signals as well as the telescope response versus the frequency. These include the galactic background temperature ($T_{\rm{sky}} \propto f^{-2.5\pm0.1}$), the power of RFI signals, particularly from broadcasting stations ($P_{\rm{noise}} \propto f^{ -2\pm2}$ on the average in the 8-25 MHz range), and losses in the phasing systems and transmission cables ($A_{\rm{noise}} \propto f^{-0.5}$).

RFI may be particularly intense at LF. For example, in a 3 kHz band, broadcasting station signals can exceed the Galactic background by 60-80 dB, and consequently any interesting astrophysical signal by $>$ 90 dB. But as broadcasting stations have narrow bandwidths ($\leq$ 5 kHz in AM band), many spectral intervals remain ``clean'' and usable for astrophysical research down to or even below 10 MHz (see. Sect. 4). Therefore, the main hindering factor is not the interference itself, but its intermodulation components produced by non-linear response of telescope elements along the signal path. The main requirement for the system is thus to minimize the intermodulation products that might otherwise contaminate the entire frequency range. The PAS that compensate for losses (and possibly for the Galactic background spectrum) and provide sufficient signal power to the receivers generally have a dynamic range much higher and an intrinsic noise much lower than those of the receiving equipment. Besides input amplifiers or attenuators and filters, the main element responsible for the linearity and dynamic range of a digital receiver is thus the ADC. 

As the full scale dynamic range of a 16-bit ADC does not usually exceed 80 dB, one role of the PAS is also to adapt the signal level to the input voltage range of the ADC. In particular, the input signal should be $\geq$20 dB above the ADC noise at any frequency in the band of interest in order to limit the ADC intrinsic noise to $\leq$1\%. If RFI does not exceed the Galactic background by more than 60 dB, a full scale dynamic range of 80 dB is sufficient to avoid most of the intermodulation associated with ADC overflow. To bring this overflow issue down to zero $\geq$20-bit ADCs are required, but those still have a limited bandwidth.

Actually, receivers operating in relatively narrow frequency bands where RFI does not exceed 30$\pm$10 dB above the background may be satisfied with a lower dynamic range. An example is the Portable Pulsar Receiver\footnote{http://www.rri.res.in/}, which works fine with a 2-bit ADC at the GEETEE radio telescope in Gauribidanur, because its bandwidth is $\sim$2 MHz at a central frequency of 34 MHz. LOFAR's LBA operate correctly with 12-bit ADC, probably due to the LBA antenna response that strongly peaks at $\sim$55 MHz, where RFI are more rare than at lower frequencies, so that the impact of antenna response and RFI do not add up to require a higher dynamic range. 16 bit is the minimum value needed for operation in the 20-88 MHz range of the LWA. But when operating in the 8-32 MHz (UTR-2, URAN) or 8-80 MHz (GURT) ranges, 16-bit ADC may not be sufficient. Therefore, low-pass or band-pass filters (16-33 MHz for UTR-2, and 25-80 MHz for GURT) are presently used during daytime observations when the RFI situation is very bad, to cut off the most polluted parts of the spectral range.

To exploit at best the ``clean'' areas of the LF spectrum not affected by broadcasting stations, the spectral resolution (channel width) of a digital receiver should be not broader than the typical RFI signal bandwidth. But generation of very narrow channels requires computing the Fourier transform over a long time window, during which broadband RFI can pollute the observation, so that ``too'' narrow channels should also be avoided. Ideally, the temporal and spectral resolutions of the receiver should be selectable. Our large archive of past LF observations suggests that optimum resolutions are 1 kHz $\leq \delta f \leq$ 4 kHz and 0.5 ms $\leq \delta t \leq$ 2 ms. An ideal observation method consists in recording waveforms (continuous flows of raw ADC data) that can be Fourier transformed offline with arbitrary temporal and spectral resolutions (with the only limitation $\delta f \times \delta t \geq 1$) depending on the astrophysical target and observing conditions, but this method requires massive data storage space. Thus it should be used only for  high resolutions measurements of specific astrophysical interest.

As a summary, digital receivers for LF radio astronomy should have the following characteristics:\\
1) high linearity and large number of bits of the ADC for ensuring low ADC noise and preventing overflow;\\ 
2) high temporal and spectral resolutions, that can be selected according to the astrophysical target and RFI conditions;\\
3) no data loss during long observation sessions;\\
4) flexible operation modes and possibility to record waveform data to disk or process it on-the-fly.

% Even provided an ideal alignment of the galactic background noise level in the entire frequency range, in its lower part interference may be 60-90 dB above the signal within a few kHz band. It may require additional efforts to correct the transfer characteristic of the amplification system. To conclude, digital receivers for low-frequency radio astronomy must have the following features: (i) have no data loss in course of long observation sessions, (ii) have a large dynamic range and small self-noise. Different operation modes should allow recording data to a disk or process it on-line.

\subsection{Previous generations of digital receivers with frequency conversion}    %sect 2-3

Since the late 1970s 1-bit correlometers were successfully used at UTR-2 \cite{Konovalenko_1982_Spectr_weak_lines}. They proved efficient for studying radio sources displaying spectral lines, e.g. radio recombination lines. Carbon radio recombination lines were discovered \cite{Konovalenko1981} and studied at the lowest frequencies \cite{Stepkin2007} using these correlometers. But their limited sampling rate of input signals prevented their use for studying astrophysical sources with broad spectra. Multi-bit high-speed digital receivers capable of continuous data recording were used at UTR-2 since 1998. These receivers performed on-line Fast Fourier Transform (FFT) or recorded raw waveforms. The first devices of this type had a limited bandwidth ($\sim$10 MHz), therefore frequency conversion was necessary. As analog mixers, even having a relatively high dynamic range, are essentially non-linear devices, they limited the linearity of these receivers. On another hand, their relatively narrow frequency band centered on a tunable frequency permitted to avoid the most polluted regions of the decameter spectrum. Table~\ref{tbl2} lists the main characteristics of the heterodyne receivers used at UTR-2 and URAN.

\begin{wstable}[h]												% Table 2
\caption{Parameters of digital receivers with frequency conversion.}
\begin{tabular}{@{}ccccc@{}} \toprule
Receiver & Digital spectro-  & ROBIN-2\tnote{b} & Portable Pulsar  & Waveform receiver,  \\
 & polarimeter\tnote{a} &  & Reseiver\tnote{c} & WFR\tnote{d}  \\ \colrule
Operating mode\hphantom{00} & \hphantom{0} Online FFT & \hphantom{0}Online FFT & Waveform & Waveform\\
FFT chan. number\hphantom{00} & \hphantom{0} 1024 & \hphantom{0}256-2048 & - & - \\
Number of inputs\hphantom{00} & \hphantom{0} 2 & \hphantom{0}2 & 2 & 2 \\
Number of ADC bits\hphantom{00} & \hphantom{0} 12 & \hphantom{0}12 & 2 & 8 \\
$f_{\rm{sampling}}$ (MHz)\hphantom{00} & \hphantom{0} 25 & \hphantom{0}30 & 3 & 6 / 14.6\\ 
Bandwidth (MHz)\hphantom{00} & \hphantom{0} 12.5 & \hphantom{0}14 & 1.5 & 3 / 7.3 \\
Temporal resolution\hphantom{00} & \hphantom{0} 2 ms & \hphantom{0}1 ms & 330 ns & 170 / 70 ns\\
Frequency resolution (kHz)\hphantom{00} & \hphantom{0} 12 & \hphantom{0}7 & - & - \\
 \botrule
\end{tabular}
\begin{tablenotes}
\item[a] \citet{Kleewein1996}
\item[b] \citet{Rucker2002}
\item[c] \citet{rri-ppr}
\item[d] \citet{Zakharenko2007}
\end{tablenotes}
\label{tbl2}
\end{wstable}

Many astrophysical phenomena and types of radio sources were studied using these receivers. The Digital spectro-polarimeter was used for a broadband study of Solar and Jovian radio emissions \cite{Konovalenko2001} and helped to discover the anomalously intense pulses of pulsars in the LF range \cite{Ulyanov2006,Ulyanov2007}. The Portable Pulsar Receiver permitted the detection of giant pulses of the pulsar in the Crab Nebula at 23 MHz \cite{Popov2006}. The fine temporal structure of lightning in Saturn's atmosphere was first studies using the WFR \cite{Mylostna2013}.

\subsection{Baseband digital receivers for UTR-2 and URAN}   %sect 2-4

Receivers with a sampling frequency at least twice larger than the maximum frequency of operation of a telescope (32 MHz for UTR-2) can record the entire working band without frequency conversion, and are thus called baseband receivers. Their advantage is their broad bandwidth and the absence of a nonlinear mixer that allows to achieve a higher dynamic range of the entire receiving system.

The first baseband receiver that started operation at UTR-2, in 2006 (named DSPZ for ``Digital Spectro-Polarimeter, type Z'') was designed specifically for decameter radio astronomical observations. Its structure and operation principles are described in detail in \citet{2007msw..conf..736K,2010A&A...510A..16R}. It permitted further (broadband) studies of Jupiter, Saturn's lightning, radio recombination lines, pulsars, searches for radio signals from flare stars and exoplanets \dots Its main parameters are listed in Table~\ref{tbl3}.

\begin{wstable}[h]												% Table 3
	\caption{Parameters of the digital baseband receiver DSPZ.}
	\begin{tabular}{@{}cc@{}} \toprule
		Number of input channels & 2 \\
		Frequency band & 0-35 MHz\\
		Sampling frequency & internal clock: 66 MHz / external: 10-70 MHz\\
		Time synchronization & 1 pps input \\
		Fast Fourier transform & Auto- and complex cross-correlation spectra \\
		FFT window length / bits of resolution & 16 384 / 30 \\
		Number of frequency channels & 8 192 \\
		Frequency resolution at $ f_{\rm{clock}} $ = 66 MHz & 4.0 kHz \\
		Integration time at $ f_{\rm{clock}} $ = 66 MHz & 0.25-128 ms\\
		ADC resolution & 16 bits \\
		Waveform recording & of unlimited duration \\
		Input impedance & 50 Ohm\\
		Input signal amplitude (max) & $\pm$ 1 V\\
		SNR, dB Full Scale & 74 dB\\
		Noise floor, dB Full Scale & $ - $117 dB\\
		Time window shape & User-defined (default = Hanning)\\
		FFT window overlap & 50\% \\
		Accumulated values precision & 54 bits\\
		\botrule
	\end{tabular}
	\label{tbl3}
\end{wstable}

\section{New digital radio astronomy receivers}    %sect 3

With the development of GURT, we needed a digital receiver with characteristics improved from DSPZ, e.g. a higher sampling frequency, more flexibility in the choice of the spectral resolution, or the possibility to synchronize several receivers at the ADC sampling period. Thus we developed the Advanced Digital Receiver (ADR) and in parallel we upgraded the DSPZ at UTR-2 and URAN radio telescopes.

\subsection{DSPZ upgrade at UTR-2 and URAN} %sect 3-1

As said in Section 2.1, UTR-2 can synthesize five simultaneous beams in the North-South direction. These can increase survey speed, or provide a simultaneous On / Off mode allowing us to distinguish weak cosmic signals from RFI. 
Optimal exploitation of this 5-beam mode benefits from synchronization of the digital signals from each beam at the precision of a single ADC sample. This capability was implemented in DSPZ in 2010 by the addition of four complementary counters to the data flow:

- second since the start of the day,

- number of frame ($f_{clock}$ sample) since the beginning of the current second (i.e. phase of the current second),

- number of spectrum or waveform block (corresponding to 8190 samples from each of the 2 input channels, to which counters are added so that data blocks have a size multiple of 8192 bytes),

- microsecond since the start of observation.\\
These counters are used for integrity check and synchronization of the data from different DSPZ receivers operating in parallel. The first two counters are placed inside the FPGA block around the ADC, that operates independently of the data processing system.
Absolute time and reliability of the time scale during long-term recordings is obtained via an external clock (frequency synthesizer) synchronized on a rubidium standard clock. 

\begin{figure}[h]												% Fig_DSP-Z_5
\begin{center}
\includegraphics[width=0.8\linewidth]{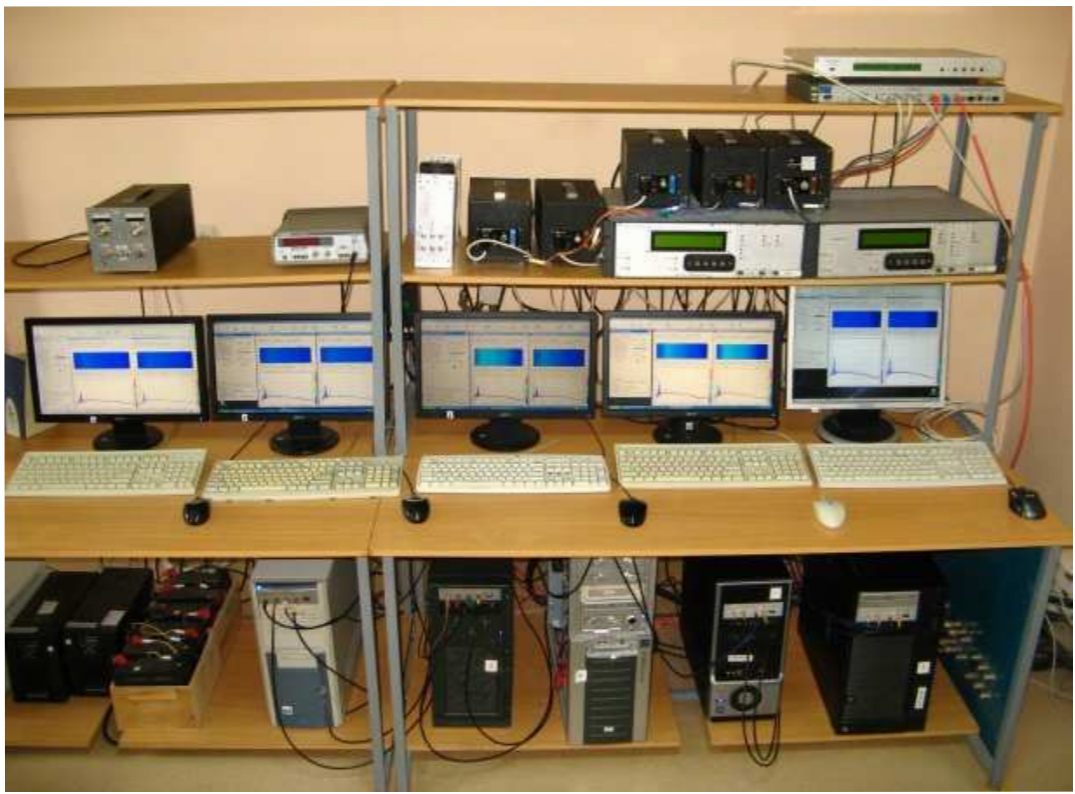}  
\end{center}
\caption{The set of five DSPZ (one per telescope beam) installed in the UTR-2 receivers room.}
\label{DSP-Z_5}
\end{figure}

Early DSPZ versions only returned two parallel data streams computed from the 2 input channels: auto-correlations of each of the 2 inputs, or auto-correlation of 1 selected input and amplitude of the cross-correlation spectrum of the 2 inputs. At the occasion of the 2010 upgrade, the capability was added for the software to deliver the full complex cross-correlation spectrum (sine and cosine parts of the cross spectrum) in addition to the auto-correlation spectra of the 2 input channels, i.e. four parallel data streams. This allows us to perform full polarization measurements at the URAN telescopes (where dual-polarized antennas are available), computing the 4 Stokes parameters from the 4 above data outputs. At UTR-2 (where antennas are single-polarized), it provides the possibility to compute the product of EW and NS antenna signals, synthesizing directly the pencil beam of 0.5$^\circ$ diameter that results from the product of the beams of the two arms.

The DSPZ already had a large input and ADC dynamic range, well adapted to the background variations in its broad spectral range and capable of operating in harsh RFI conditions (see Fig.~\ref{eDay}). With its new capabilities, the upgraded DSPZ meets most technical requirements of ideal LF radio astronomy receivers. 
It is the main type of receiver installed at UTR-2 (one per beam output -- Fig. 5) and at the URAN-2 and -3 radio telescopes. The possibility to compute full auto- and cross-spectra on-the-fly, and to record data synchronously with all DSPZ receivers provided new multi-beam and interferometric capabilities. Those considerably widened the variety of astrophysical studies carried on with these telescopes. In Section 4 we show examples of recently obtained scientific results. 

\subsection{Digital receiver for GURT}

The ADR developed for GURT has a higher sampling frequency and thus a broader spectral range of operation.
The design of this receiver is built on all the technical advances from the DSPZ, but also includes new capabilities: addition/subtraction mode, digital signal delay between its two input channels, client-server architecture of its control software.

GURT is composed of many identical subarrays, each being internally phased using analog phase shifters, but the subsequent phasing of subarrays relative to each other will be digital. The maximum ADC sampling rate was extended to 160 MHz\footnote{the LTC2209 circuit is used} to meet the 8-80 MHz range of operation of GURT.

The ADR has a modular design, composed of an ADC board, a FPGA module, and a high-speed (10 Gb/s Ethernet) waveform data transfer module. This enables various block combinations. A receiver with all 3 blocks can operate in both spectral (on-the-fly FFT) and waveform capture modes. A receiver without the 10 Gb/s Ethernet module operates only in spectral mode (up to 16384 frequency channels, with a temporal resolution down to 3 ms). A receiver without the FPGA module provides the digital waveforms from the 2 polarized outputs of a subarray, that can be injected into the digital phasing system of GURT. The latter set (ADC + 10 Gb/s Ethernet module) is planned to be used at each GURT subarray, spectral calculations being performed after the digital phasing (beamforming) of several subarrays, using one receiver (with the FPGA module) per synthesized beam.

The ADR receiver can be used as a simple 2-input digital phase shifter. Phase shifting can be done in the Fourier domain, but in addition a temporal delay between signals of the two channels can be introduced. The specific circuit that builds the delay provides 768 delay steps per sampling period, corresponding to 8 ps steps for $f_{clock}$ = 160 MHz. The maximum delay between the two channels is 128 sampling periods, i.e. up to 800 ns. With a temporal delay ranging from 8 ps to 0.8 $\mu$s, the phasing of two subarrays spaced by 240 m can be achieved with 2.4 mm precision. 

The ADR receiver allows to independently adjust input signal amplitudes (e.g. for normalization purposes) by multiplication by a coefficient k=n/8192 (with 1 $ \leq $ n $ < $ 65536). 

At UTR-2, before the availability of the upgraded DSPZ, receivers (analog or digital) generally provided only the auto-correlations of the input signals. It was thus imagined to form with analog circuits (summators, broadband $\pi$ phase shifter) the sum and the difference of the 2 input signals, that were then injected in the receiver, computing the auto-correlation (or power) of each input. The ``sum'' power ($\Sigma=(U_1+U_2)^2$) corresponds to a beam in the sky with the shape of a large ($\sim10^\circ$) cross, with an enhanced ($\times4$) pencil beam at the center. The ``difference'' power ($\Delta=(U_1-U_2)^2$) corresponds to the same beam as $\Sigma$, but with the pencil beam set to zero.The difference of these 2 outputs is proportional to the product of the 2 initial antenna signals, according to the expression:

\begin{equation}										%equation 1
\Sigma - \Delta = (U_1+U_2)^2 - (U_1-U_2)^2 = 4 \cdot U_1U_2~ ,
\label{mult}
\end{equation}

that corresponds to the pencil beam $P$ only. These outputs $\Sigma$, $\Delta$, and $P$ had several astronomical applications, especially at the T-shaped UTR-2: they allowed us to form a 0.5$^\circ$ pencil beam ($P$), much less susceptible to RFI than the elongated EW and NS beams of $\sim 0.5^\circ \times 10^\circ$. With two or more receivers, and using the same number of different beams in the North-South direction, several pencil beams can be synthesized, enabling e.g. simultaneous On / Off observations. With a single receiver, the sum or the pencil beam ($\Sigma$ or $P$) to the target provides the On signal whereas the difference ($\Delta$) provides a simultaneous Off signal \cite{vasylieva:tel-01246634}. When only the pencil beam power is required by the scientific program (e.g. the UTR-2 pulsar and transients survey \cite{2015OAP....28..252Z}), the $\Sigma-\Delta$ mode produces half the data volume than the full correlation mode.

It was accordingly decided to include in the ADR software the possibility to deliver $\Sigma$ and $\Delta$ outputs after phase shifting described above, allowing for coherent summation or difference. One ADR unit will thus be able to digitally phase-shift, coherently sum (or subtract), and accumulate (if requested) any two input signals, and output the corresponding $\Sigma$ and $\Delta$, in spectrum or waveform mode. The pencil beam signal $P=\Sigma-\Delta$ can be derived from these outputs. This mode will have many applications, in GURT (for any pair of sections or groups of sections -- GURT may be developed with a overall T-shape distribution of its sections, following the positive experience with UTR-2) as well as at UTR-2 (see e.g. section 4.3).

\begin{figure}[h]												% Fig_Block_ADR
\begin{center}
\includegraphics[width=1\linewidth]{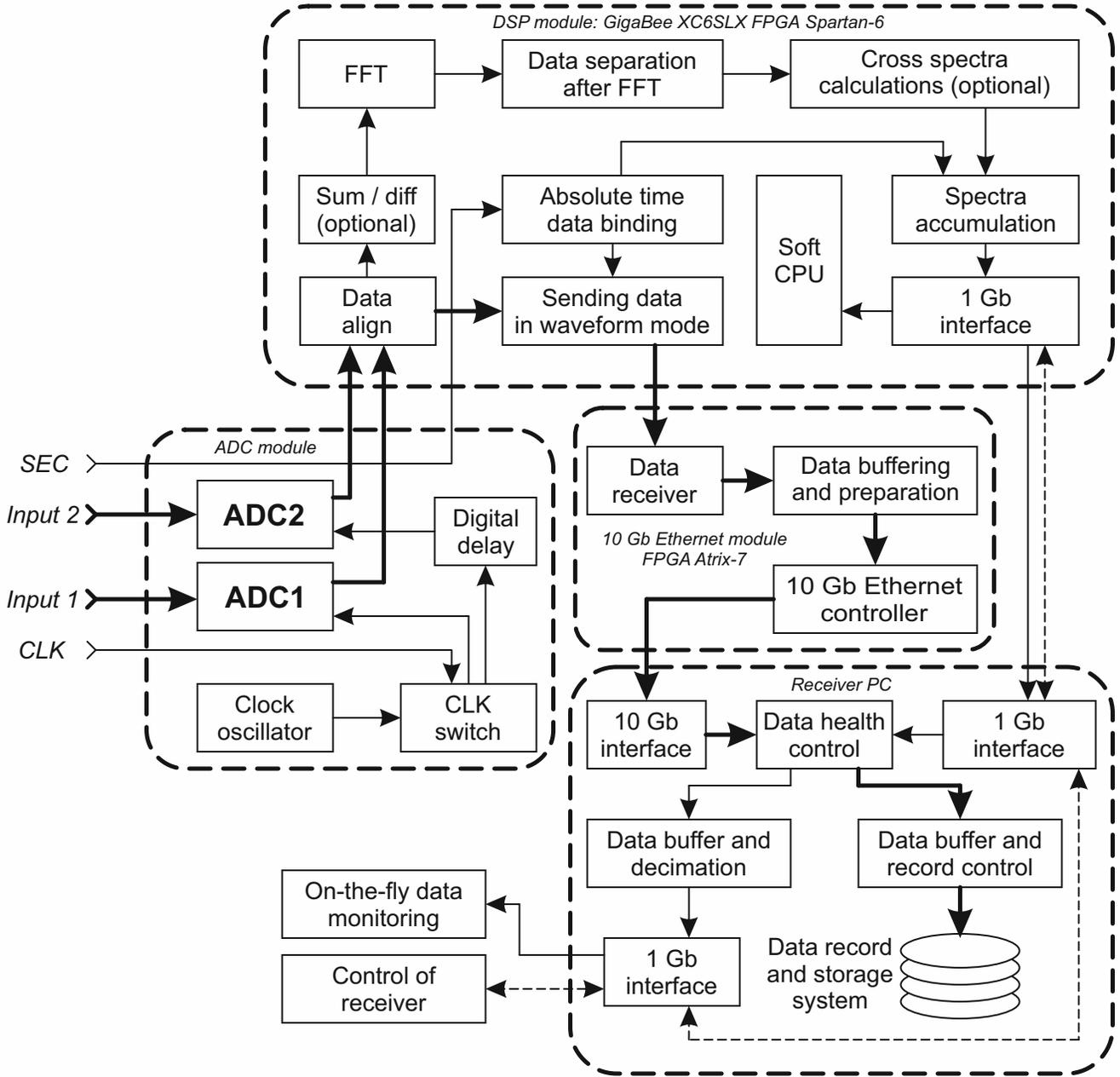}  
\end{center}
\caption{ADR System Block Diagram.}
\label{fig_Block_ADR}
\end{figure}

A block diagram of ADR digital receiver with interfaces is shown in Fig.~\ref{fig_Block_ADR}. The ADR has four inputs:
 
- "CLK" for connecting an external synchronization clock. 

- "SEC" for input of a GPS pulse per second (PPS) signal for absolute time referencing of ADC samples. 

- input 1 "A" and input 2 "B" for the analog sky signals, with 180 MHz input bandwidth. 

To calculate the spectra a standard Fast Fourier Transform algorithm is used on-the-fly:

\begin{equation}
S_{1,2}(m,k) = \sum^{N_{\rm{FFT}}-1}_{n=0} S_{1,2}(n) W(n-m) e^{-i \frac{2 \pi}{N} kn}
\label{fft}
\end{equation}
where $S_{1,2}(n)$ is the digitized signal at ADC outputs, $W(n)$ the FFT window function, $N_{\rm{FFT}} = 2^{10+k}$ with $k = 1$ to 5  the length of a data segment. Before calculating the FFT, the data is multiplied by 18-bit coefficients of a window function of same length as the data segment. In order to save FPGA resources, a custom method involving two real sequences based on a single complex FFT is used \cite{Vasilyev2014}. This algorithm requires further data arrangement into channels, but the gain in processing time is much higher than the computational burden needed for data arrangement. The FFT module processes on-the-fly the input raw data stream at the sampling frequency, without any data loss. The output data are arranged into channels as follows: 

\begin{eqnarray}
\begin{array}{l}
\Re [Y(k)] = -  \dfrac{\Re[S(k) + S(N_{\rm{FFT}} - k)]}{2},\\ [9pt] 
\Im [Y(k)] =  ~~ \dfrac{\Im[S(k) - S(N_{\rm{FFT}} - k)]}{2},\\[9pt] 
\Re [Z(k)] = ~~  \dfrac{\Im[S(k) + S(N_{\rm{FFT}} - k)]}{2},\\[9pt] 
\Im [Z(k)] = - \dfrac{\Re[S(k) - S(N_{\rm{FFT}} - k)]}{2} ,
\end{array}
\label{reform}
\end{eqnarray}
where $ S(k) $ is a complex sequence of $N_{\rm{FFT}}$ length, obtained as a result of Fourier transform (\ref{fft}) from two real inut sequences, and real $ \Re $ and imaginary $ \Im $ parts of the complex sequences $ Y(k) $ and $ Z(k) $ (for $ k = 1, ...  N_{\rm{FFT}}/2$) are the output spectra from the two real input channels. For zero frequency, formulas (\ref{reform}) are not correct, but this is acceptable because the DC component value in the spectra is due to the ADC input bias, which does not carry any useful information.

The power spectra of both channels $ P1 $ and $ P2 $, as well as  the real and imaginary parts of the cross-spectra are calculated according to the formulas:

\begin{eqnarray}
\begin{array}{l}
P1(k) = \Re^2[Y(k)] + \Im^2[Y(k)]\\ 
P2(k) = \Re^2[Z(k)] + \Im^2[Z(k)]
\end{array},
\label{p1p2}
\end{eqnarray}

\begin{eqnarray}
\begin{array}{l}
\Re [k] = \Re[Y(k)]\Re [Z(k)] + \Im [Y(k)]\Im [Z(k)]\\
\Im [k] = \Im [Y(k)]\Re [Z(k)] - \Re[Y(k)]\Im [Z(k)]
\end{array}.
\label{ReIm}
\end{eqnarray}

\begin{figure}[h]												% Fig Noise_ADR. fig8
\begin{center}
\includegraphics[width=0.5\linewidth]{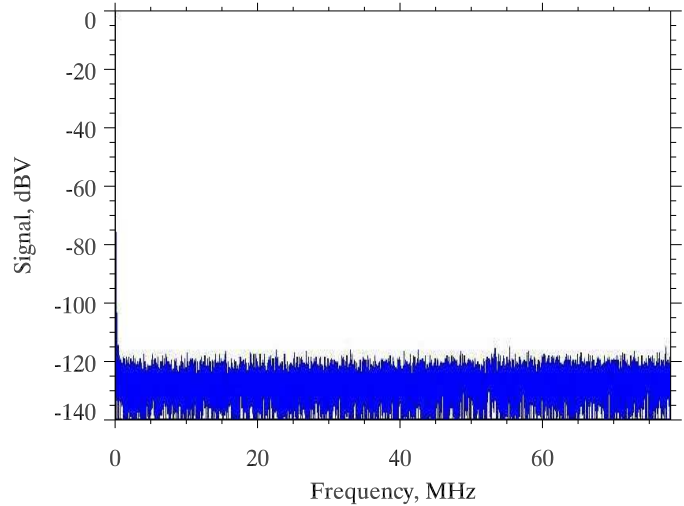}  %100 percent
\end{center}
\caption{Spectrum of the ADR self-noise at the highest spectral resolution (16384 channels, $f_{\rm{clock}}$ = 156 MHz). 
% 156 of 160 MHz ? (as below)
In spite of the fact that the FPGA and the fast data transfer modules are potential strong digital RFI sources, the hardware and software (FFT computation and averaging) design of the ADR results in no spurious spectral component emerging from the noise level.}
\label{fig_foto3}
\end{figure}

The power spectra samples (\ref{p1p2}) and the complex cross-spectra samples (\ref{ReIm}) are then averaged in the FPGA module. The maximum number of time integrations $ N_{\rm{AVR}} $ in the ADR is 32 times larger than that in the DSPZ. This corresponds to a SNR increase of 7.5 dB for a stationary signal occupying a single spectral channel. Real and imaginary parts of cross-spectra (\ref{ReIm}) are averaged separately. The averaged data is stored in a special format, with 30-bit mantissa and 6-bit exponent. We developed this format taking into account the computational resources of the FPGA board and the characteristics of signals to be processed. The resulting data is transferred to a PC with a truncated 26-bit mantissa. This optimizes the use of the memory module memory while at the same time preserving data accuracy. The relative rounding error $\varepsilon$  for the implemented averaging algorithm is $\varepsilon = 0.5 \times 2^{1-m}$, with $m$ the number of mantissa digits. It is equal to $9.31 \times 10^{-10}$ for the 30-bit mantissa and $1.49 \times 10^{-8}$ with the truncated 26-bit mantissa,
% not 2.98 \times 10^{-8}$  :   0.5*2^(1.-26) = 1.49012e-08
20 dB below the receiver noise, thus without any effect on the measured signals (Fig.~\ref{fig_foto3} ). The averaging algorithm implemented does not induce any spurious spectral components due to rounding.

\begin{figure}[h]												% Fig FPGA fig7
\begin{center}
\includegraphics[width=0.5\linewidth]{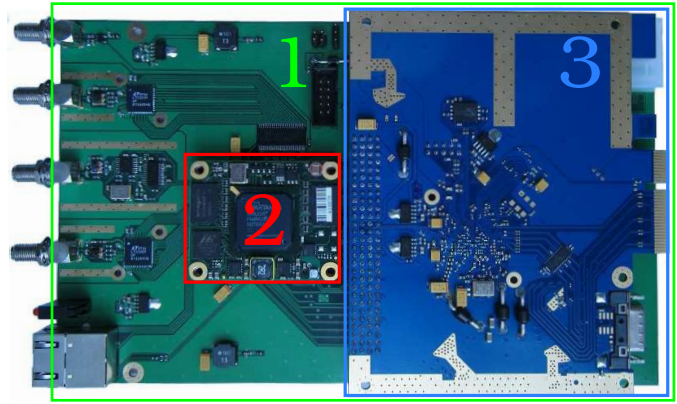}  %100 percent
\end{center}
\caption{ADC board (1), FPGA module (2) and high-speed waveform data transfer module (3).}
\label{fig_FPGA}
\end{figure}

The ADR system software has a client-server architecture. The ADR Server runs on the host PC. It communicates with the ADR DSP-board via a 1Gb Ethernet interface (that sends control commands and receives the responses). The ADR Server receives data streams in Spectrometer and Waveform modes via the 1 Gb and 10 Gb Ethernet interfaces, respectively. 
As mentioned above, the ADC sampling clock signal in the second channel has a built-in controlled digital delay (providing 768 delay steps per sampling period) implemented on a CDCF5801A chip. Data from both channels are thus sampled at the same frequency but with a controlled phase difference. 
The 10 Gb Ethernet module specially developed for the ADR provides lossless transfer of digitized data at 640 Mbytes/s (with 160 MHz sampling frequency $\times$ 2 inputs channels $\times$ 16 bits). The module is assembled on a separate board (see Fig.~\ref{fig_FPGA}) using Xilinxâ„¢ Artix-7 FPGA. 
Data processing and receiver control are implemented on a GigaBee XC6SLX board based on a Xilinx$\texttrademark$ Spartan-6 FPGA (Fig.~\ref{fig_FPGA}). 
Data from the ADC is transferred through the Spartan 6 FPGA to the 10 Gb Ethernet board using an LVDS interface. To reduce the influence of RFI picked up by the highly sensitive ADC inputs and minimize the number of connection lines, the data are multiplexed with a factor of 4, and in this way the bit rate of each of the 9 LVDS data lines is 640 Mb/s. 
The ADR server has a 10 Gb network adapter Chelsio S320E-CXA, and the UDP protocol with 8 kB datagram length ("jumbo-frame") is used for data transfer. 
The DSP board includes 128 Mb DDR3 RAM for data buffering. Data recording and memory reading is performed with 128 bit words, resulting in a data rate of $128/8 \times 160$ MHz = 2.56 Gbyte/s. A FIFO (first in first out) buffer is used to reduce the peak load, resulting in an average data rate of 520 Mbyte/s, thus below the DDR3 limit rate of 650 Mbyte/s.
The local storage system is based on a hardware RAID controller LSI SAS 9211-8i (RAID 0 configuration).
% I tried to put the above paragraphs in logical order (it was all mixed). Ideally, the description should follow the data path in the receiver. Check if it is ok with you, and modify if not.

The same counters of seconds, frame number (phase of the current second) and spectrum number as for the DSPZ receiver are used for data integrity check in the ADR. Referencing of every sample to absolute time ensures precise temporal synchronization of data from different receivers, which necessary for observations in VLBI mode, aperture synthesis, and joint processing of data recorded simultaneously with several receivers such as ADR and DSPZ. These characteristics gives the unique possibility to add together UTR-2 and GURT effective areas by means of digital phasing of waveform data obtained simultaneously by UTR-2 and the GURT subarrays.

\begin{figure}[h]												% Fig_foto1 fig9
\begin{center}
\includegraphics[width=0.5\linewidth]{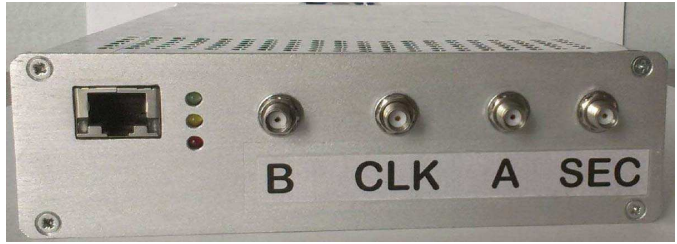} 
\end{center}
\caption{Front view of the ADR box, that contains the boards displayed in Fig.~\ref{fig_FPGA}. The input labels correspond to those of Fig.~\ref{fig_Block_ADR} (with for $A$= input 1 and $B$= input 2). The receiver can operate independently in a section of the radio telescope or installed in the ADR Server).}
% last sentence not clear
\label{fig_foto1}
\end{figure}

All modules of an ADR are assembled in a single box with "5.25 inch half-height" form factor, whose dimensions are 5.75 in $ \times $ 8 in $ \times $ 1.63 in (146.1 mm $ \times $ 203 mm $ \times $ 41.4 mm). The front view of the ADR is displayed on Fig.~\ref{fig_foto1}. The ADR box is mounted into a 5.75 inch slot of its host (desktop) PC.

Parameters of the ADR receiver are listed in Table~\ref{tbl4}.

\begin{wstable}[h]												% Table 4
\caption{Parameters of digital baseband receiver DSPZ.}
\begin{tabular}{@{}cc@{}} \toprule
Number of input channels & 2 \\ [3pt]
Analog input bandwidth & 180 MHz \\ [3pt]
Input impedance & 50 Ohm \\ [3pt]
Input voltage & Â±1 V \\ [3pt]
ADC sampling frequency  & internal: 156 MHz / external: 20-160 MHz \\ [3pt]
ADC resolution & 16 bits \\ [3pt]
ADC intrinsic dynamic range & 73 dB \\ [3pt]
SFDR (spurious free dynamic range) \\ (16384 samples per FFT) & 112 dB \\ [3pt]
Intrinsic noise level \\ (16384 samples per FFT) & -117 dB \\ [3pt]
Digital DC bias compensation & No \\ [3pt]
Dithering option \\ (for an increasing SFDR value) & Yes \\ [3pt]
FFT size (samples or spectral channels) & 2048, 4096, 8192, 16384, and 32768 \\ [3pt]
Output FFT samples resolution & 32 bit \\ [3pt]
Speed of processing & 4800 complex 32768 points FFT per second \\ [3pt]
Count of averaged spectra & 16 - 32768 \\ [3pt]
Selectable frequency band output & by groups of 1024 spectral channels \\ [6pt]
 "Spectrometer" sub-modes & 
 \begin{tabular}{c}
 1. â€œAâ€  channel spectrum output \\
 2. â€œBâ€  channel spectrum output \\
 3. â€œA$\&$Bâ€  channels spectrum output \\
 4. â€œA+Bâ€ $\&$ â€œA-Bâ€ channels spectrum output \\
 5. â€œA$\&$Bâ€  channels spectrum and \\ 
 cross-correlation between â€œAâ€ $\&$ â€œBâ€ channels spectra\\ [6pt]
 \end{tabular}
 \\ 
â€œWaveformâ€ sub-modes &
\begin{tabular}{c}
1. â€œAâ€  channel waveform output \\
2. â€œBâ€  channel waveform output \\
3. â€œA$\&$Bâ€ channels waveform output \\
4. â€œA+Bâ€ $\&$ â€œA-Bâ€ channels waveform output\\ [6pt]
\end{tabular}\\ [3pt]
Output type (for â€œSpectrometerâ€ mode) & 1 Gb Ethernet \\ [3pt]
Connection cables (for â€œSpectrometerâ€ mode) & Cat.5 UTP \\ [3pt]
Output type (for â€œWaveformâ€ mode) & 10 Gb Ethernet \\ [3pt]
Maximal data rate to host PC (â€œWaveformâ€ mode) & 650 MB/s \\ [3pt]
Maximal data rate to host PC (â€œSpectrometerâ€ mode) & 80 MB/sec \\ [3pt]
Control interface & TCP/IP \\ [3pt]
Data interface & UDP/IP \\ 
\botrule
\end{tabular}
\label{tbl4}
\end{wstable}

\section{Assets of new generation digital receivers and applications to low-frequency radio astronomy}

Evidently the essential improvement of receiver parameters (widening of bandwidth, increasing of temporal and spectral resolution, increasing of operational modes of ADR) allows conducting more various, reliable and high-quality radio astronomical studies. The same applies to DSPZ upgrade which was intended to provide means for synchronous operation of several receivers. Results of processing the data recorded with the developed receivers in numerous astronomical observations on a variety of research programs confirm their high quality. 
Stability of parameters, high dynamic range and sensitivity yield the absence of intermodulations and overflows under bad interference conditions, and increase the SNR during long-term accumulation of the pulsar pulses, spectral lines, etc. Observation technique can be adapted to various space objects of interest, due to a variety of operating modes. This increases the reliability of the results. The examples of some observation data that illustrate the advantages of digital receivers are given below.

% Immunity to RFI
\subsection{Immunity to RFI}

High interference immunity of digital receivers combined with a large effective area of the LF UTR-2 and URAN radio telescopes, or with a low noise and broad band of GURT subarrays can sustain any RFI conditions, maintain operability and yield unique scientific results. 

% Immunity to RFI
%High linearity and broad band of both GURT preamplification system and ADR enable operation in the 8-80 MHz frequency range despite the impact of powerful interference of broadcasting stations.

%High-quality of observations relies not only on the high linearity of the receiver, but also on the high frequency resolution. 
Efficiency of RFI mitigation procedures in difficult observation conditions can be achieved not only with high linearity but also due to the narrow bandwidths of frequency channels in digital receivers (1-4 kHz, according to the mode and the selected sampling frequency), which is considerably smaller than or comparable to the emission bandwidth of powerful radio stations (5-10 kHz). Applying the Hanning window with a low side lobes level during the Fourier transform facilitates eliminating interference signals from the spectrum. The advanced post-processing algorithms \cite{2010MNRAS.405..155O,1997pre4.conf..101Z,2013MNRAS.431.3624Z,vasylieva:tel-01246634} perform an effective RFI mitigation by identifying the distinctive features of interference signals compared to cosmic sources.

% Pulsar observations
%In addition the using of â€œphase of secondâ€ conter allows controlling data integrity and time binding, which is very important for long-term pulsar observations. 

\begin{figure}%[h]												% Fig PSRB0809_8-33MHz
	\begin{center}
		\includegraphics[width=0.5\linewidth]{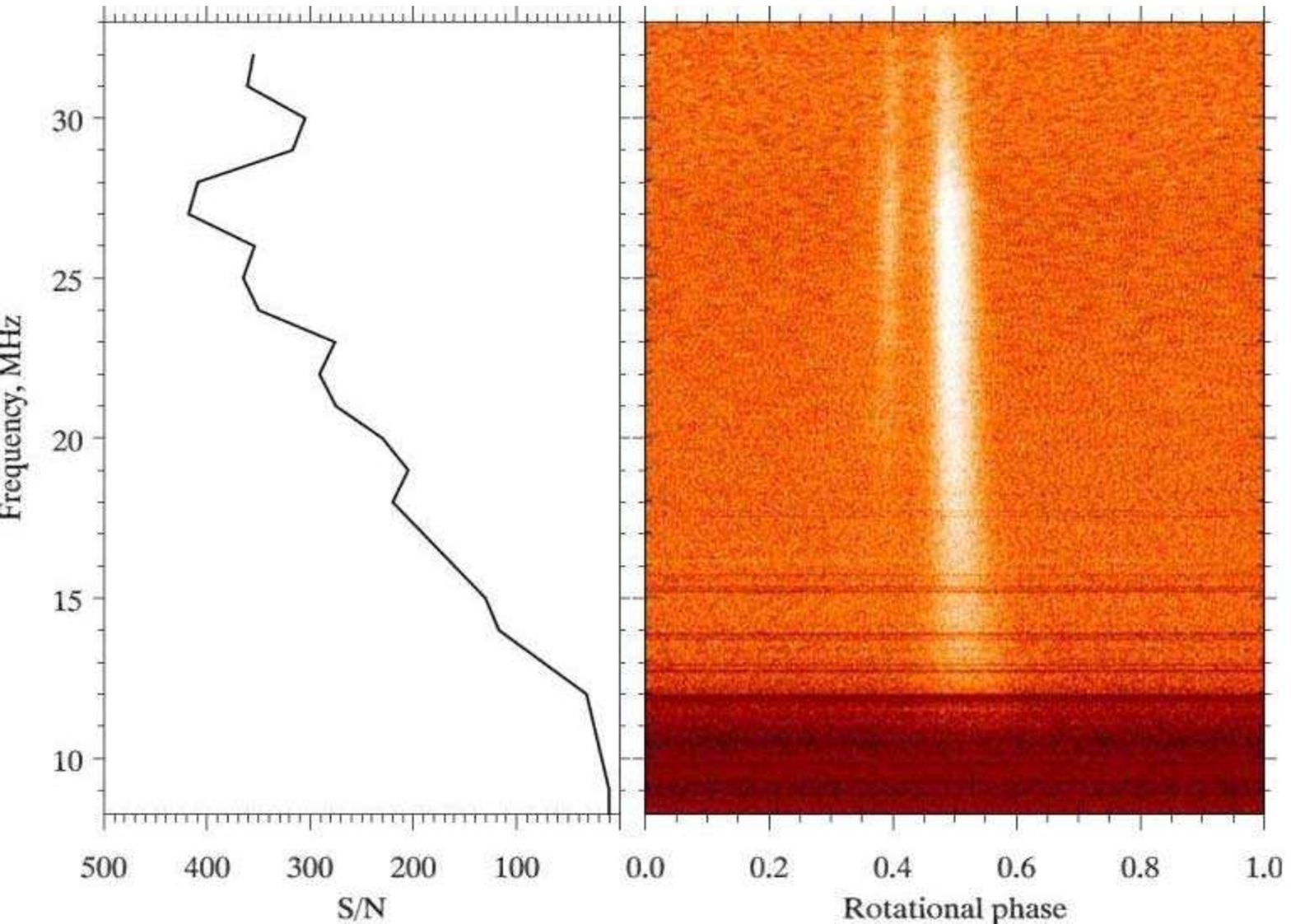} %100 percent
	\end{center}
	\caption{SNR for the average profile of the pulsar B0809+74 (1 hour integration time) at 1 MHz subbands in a range of 8 to 33 MHz (left), spectrogram of the average profile of the same pulsar (right, DSPZ receiver at UTR-2 radio telescope, observations Dec 25, 2013, start 21:14 UT).}
	\label{PSRB0809_8-33MHz}
\end{figure}

As a good example of RFI immunity Fig.~\ref{PSRB0809_8-33MHz} demonstrates a spectrogram of the average profile of the pulsar B0809+74 over the entire operating frequency range of UTR-2 (8-33 MHz, DSPZ receiver).

The left panel of Fig.~\ref{PSRB0809_8-33MHz} shows SNR in a 1 MHz partial band, after 1 hour of night-time integration. In spite of strong RFI with many spectral components in the range 8-12 MHz (which are 30-40 dB above the galactic background, and 50-60 dB above the pulsar signal as we can see from Fig.~\ref{eDay}) the pulsar SNR reaches $\approx$ 10 at 8-10 MHz, and up to 30 at 12 MHz. Such figures of SNR became possible due to high spectral and temporal resolutions, high linearity of the entire system and effective RFI mitigation algorithms. At higher frequencies, the SNR reaches 400 (27-28 MHz).

%The left panel of Fig.~\ref{PSRB0809_8-33MHz} shows SNR in a 1 MHz partial band, after 1 hour of integration in night time. As seen from Fig.~\ref{eDay}, even at this time in the 8-12 MHz range RFI is 30-40 dB above the galactic background, and 50-60 dB above the pulsar signal. The number of RFI spectral components is sufficiently large. Only due to high spectral and temporal resolutions and effective RFI mitigation algorithms the pulsar SNR reaches $\approx$ 10 at 8-10 MHz, and up to 30 at 12 MHz. At higher frequencies, the SNR reaches 400 (27-28 MHz).

\subsection{Solar observations}

% Solar observations
Solar radio emission is generated by many sources of different nature. Their study requires maximally wideband radio telescopes and receivers. None of enumerated in Subsections 2.3-2.4 receivers was not able to receive Sun radio emission in such a wide range (80 MHz) and harsh RFI conditions.

\begin{figure}%[h]												% Fig Sun_Gurt
	\begin{center}
		\includegraphics[width=0.9\linewidth]{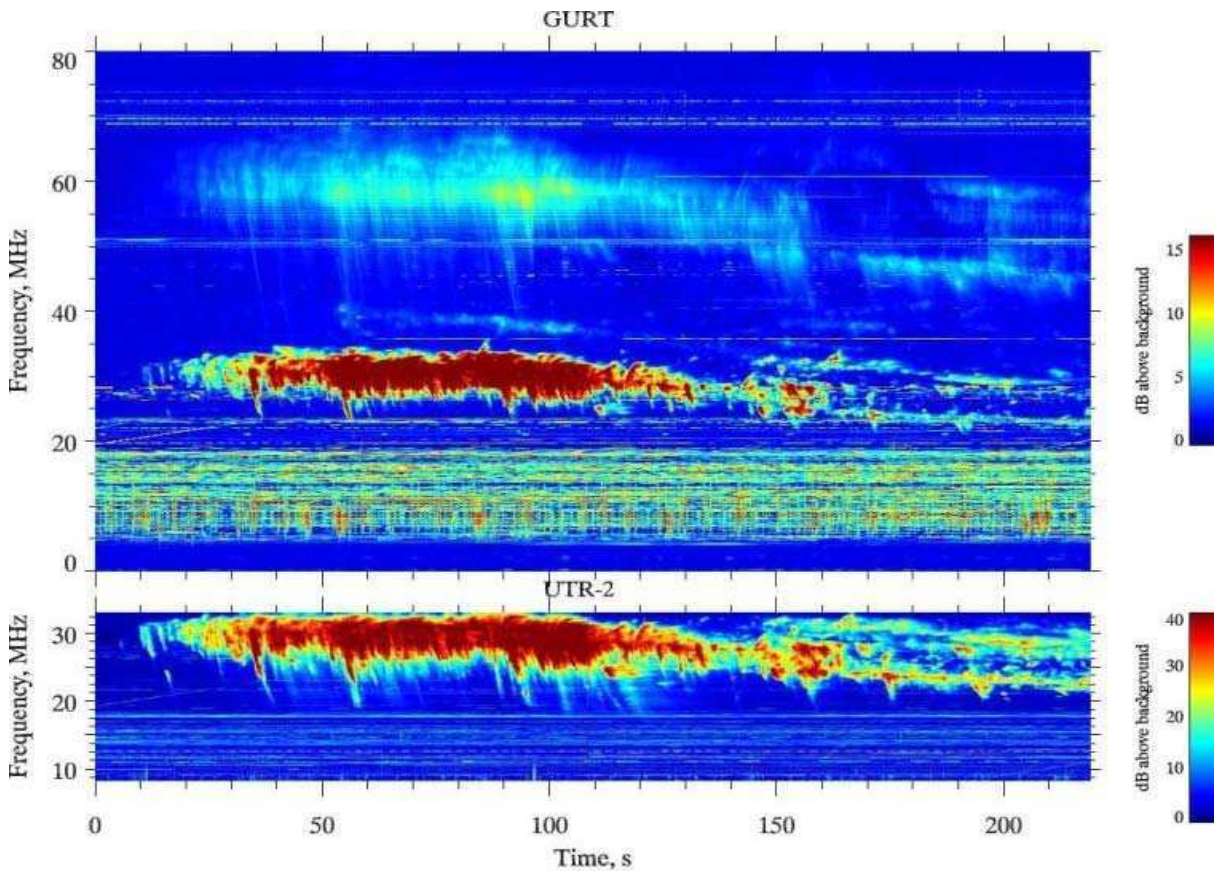}  %100 percent
	\end{center}
	\caption{Observations of a type II burst. UTR-2 frequency band is 8.25-33 MHz, frequency and time resolution of 4 kHz and 100 ms, respectively. Recording was conducted on subarray GURT in the range 8-80 MHz with the same time resolution and frequency resolution of 20 kHz. Start recording corresponds to 07:11:15 UT.}
	\label{fig_SunGURT}
\end{figure}

Fig.~\ref{fig_SunGURT} shows the result of simultaneous observations of the Solar type II burst with the above radio telescopes. The burst was recorded during observations at GURT, ADR receiver (upper panel) and UTR-2, DSPZ receiver (lower panel) on July 25th, 2014. The panels show the dynamic spectrum of 3.5 minutes length. Dynamic spectra for two linear polarizations of GURT have been summed up to improve the SNR.

The distinctive features type II bursts (about 5 min duration, slow frequency drift, presence of the second harmonic, the band splitting on both harmonics, spiky structure, etc.) are clearly seen in a wide band of the ADR receiver installed at GURT subarray. It should be noted that intense daytime RFI does not obstruct obtaining reliable astrophysical results. A broad band of RFI from 5 to 18 MHz (visible in GURT data panel) has been suppressed in the UTR-2 data due to the high spatial selection of the radio telescope and a low level of side lobes. 

% Solar observations
The next figure (Fig.~\ref{Sun_9-33_MHz}) shows potential of Solar bursts observations down to 9 MHz. Residuals from RFI mitigation procedure do not affect accurate parameter classification at extremely low frequencies.

\begin{figure}[h]												% Fig Sun_9-33_MHz
\begin{center}
\includegraphics[width=0.5\linewidth]{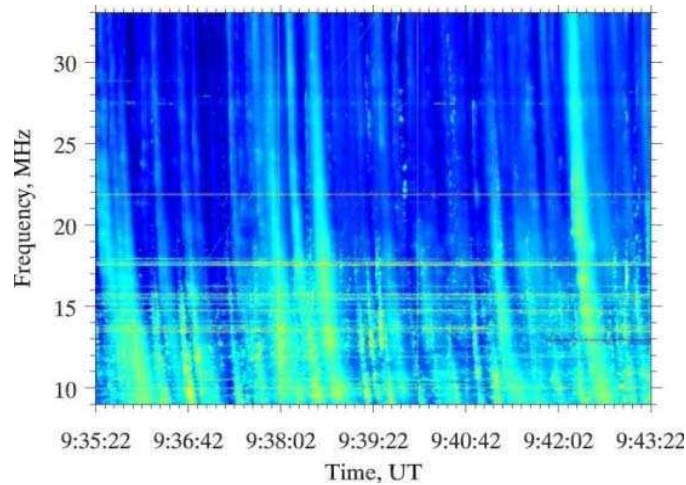} %100 percent
\end{center}
\caption{A sequence of type III Solar bursts recorded at UTR-2 in 9-33 MHz range with DSPZ receiver (observation date: July 8, 2014).}
\label{Sun_9-33_MHz}
\end{figure}

% Solar observations
In addition, the available choice of observation modes, precise time referencing, and simultaneous operation of several receivers allow us to build solar radio maps and to picture developing flares using a heliograph installed at the UTR-2 radio telescope. It also enables radio interferometric studies where separate UTR-2 sections work as spatially separated antennas.

% Saturn's lightning observations
\subsection{Saturn's lightning observations}

One of the most significant achievements of LF studies at UTR-2 was a ground-based detection of lightning in Saturnâ€™s atmosphere \cite{Zakharenko2012,Konovalenko2013} SED (Saturn Electrostatic Discharges) and the discovery of their fine temporal structure \cite{MYLOSTNA2014}. This was possible thanks to high spatial selection of UTR-2 and the use of broadband digital receivers with high temporal resolution.

In order to detect SED it is necessary to overcome a very serious challenge: to distinguish between signals of extraterrestrial lightning and strong natural and artificial discharges in the Earth's atmosphere. The most reliable way to single out the signal of interest is spatial criterion, i.e. providing an On / Off mode. Regarding the UTR-2 design there are two possible ways to organize such mode.

As UTR-2 is composed of two antennas (North-South and East-West), one of the receiver channels can be fed with the sum of these antenna signals (in single beam mode), and the other with their difference. It is a way to create an On / Off mode (see section 3.2), where the radiation pattern of the antenna sum has its maximum in the direction of Saturn (On), and the difference diagram has zero gain in this direction (Off). Side lobes of the diagrams are very similar. After subtracting (eq.~\ref{mult}) the data of the second receiver channel from the data of the first channel, interfering signals of terrestrial origin are suppressed, and the useful signal of the lightning in the Saturns atmosphere are highlighted. This approach was used during early stages of SED detection.

 Later, after the upgrade of DSPZ receiver and implementation of the correlation mode, the 5-beam mode of phased antenna array was used, when On and Off beams are formed independently and pointed at different directions. One of the receivers forms a narrow pattern (using the correlation of the North-South and East-West antenna signals) in the direction of Saturn, and the other - in the direction of the reference area of the sky,  1$^\circ$ apart to the north.

\begin{figure}[h]											% Fig SP_2016_colorbar
\begin{center}
\includegraphics[width=0.5\linewidth]{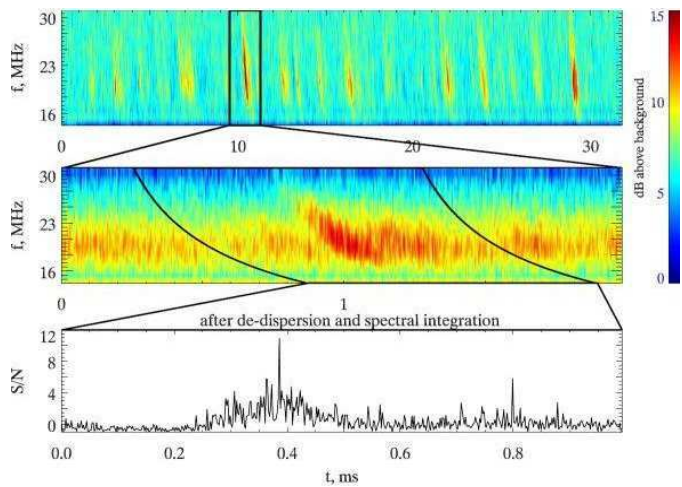} %100 percent
\end{center}
\caption{Different time scales of lightning in the atmosphere of Saturn. The sequence of bursts inside of lightning, the temporal resolution of 35 $\mu$s (upper panel). A separate splash with 8 $\mu$s time resolution (middle panel). Time profile (lower panel) after compensation of a dispersion delay in the propagation medium and accumulation over the frequency, (time resolution is 0.5~$\mu$s). Observation date Dec 23 2010, 4:00 UT}
\label{SP_2016_colorbar}
\end{figure}

As all the receivers are synchronized, the Off channel data can be correctly subtracted from the On channel, and thus the useful signal can be highlighted. Synchronous recording of the correlation and waveform data at different receivers allows (using the specially developed software) detecting the lightning signals automatically in the records of the receiver with a large integration time (10 ms, correlation mode), and find corresponding data chunks with high temporal resolution (15 ns, waveform mode). Discovery of a fine temporal structure of lightning in the Saturnâ€™s atmosphere relied entirely on the waveform mode of the DSPZ receivers. Fig.~\ref{SP_2016_colorbar} shows lightning spectrograms with different temporal resolution. The typical duration of individual discharges is 30-300 $\mu$s within the lightning that is usually 10-400 ms long (upper panel). The middle panel shows an individual burst and its temporal profile (lower panel) after dispersion delay compensation, i.e. removal of the effect of propagation medium.

%SP_2016_colorbar

\subsection{Radio Recombination Lines}% Radio Recombination Lines
\begin{figure}[h]												% Fig recombLines fig14
\begin{center}
\includegraphics[width=0.5\linewidth]{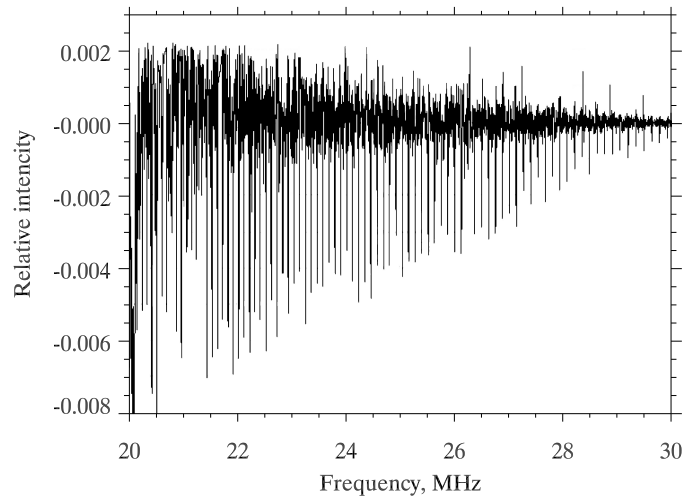} %100 percent
\end{center}
\caption{Series (about 100) of carbon radio recombination lines in the direction of Cas A in the frequency range 20-30 MHz with resolution 2 kHz. The measurements were carried out using the North-South arm of UTR-2 radio telescope after about 3 h of integration (May 12, 2012, DSPZ receiver).}
\label{recombLines}
\end{figure}

Flexible sampling rate (10-70 MHz for DSPZ and 20-160 MHz for ADR) allows increasing spectral resolution for spectral lines study. Certainly the waveform mode allows us to obtain very narrow frequency channels, but for long-term observations (10-100 hours and more) the required resources (disk space, computational burden) become unreasonably large. In the same time the change of spectral resolution can be achieved with reducing the ADC clock frequency and using additional filters. In this way it is possible to obtain a spectral resolution of 600 Hz in 10 MHz bandwidth using ADR. Fig.~\ref{recombLines} demonstrate results of radio recombination line detection in the direction towards the Cas A source achieved with signal accumulation of 3 hours. Reducing the frequency channel width (by setting $ f_{\rm{clk}} $ to 33 MHz) improves the accuracy of line profile restoration, discriminating the level of the background and the â€wingsâ€ of the line, to refine the radial velocity of the source. A broad band of the receivers is enough to accumulate and average of a large number of observed lines simultaneously, which is equivalent to increasing the integration time by the same factor. 

% Pulsar observations
\subsection{Pulsar observations}

Due to the high stability and sensitivity of the receivers, it is possible to integrate a weak signal over a long time. It is vital for studies of pulsars, as the correct integration (versus pulsar period) of the recorded signals is required to obtain an average pulse profile. 

The first redetection of Northern sky pulsars at LF \cite{2013MNRAS.431.3624Z} took place due to a large effective area and a broad frequency range of UTR-2, suitable for precise determination of pulsar dispersion measures. Precedent inaccurate knowledge of dispersion measures was the main reason why there were no previous successful detections of the above pulsars at UTR-2. 

After the commissioning of the first GURT subarray, the program for pulsar redetection was continued in a wider frequency range. This study requires that time and frequency referencing should be very precise (about 10$^{-8}$) within many hours of pulsar observations using a GURT subarray. Simultaneous observations of pulsars at UTR-2 and GURT will not only reveal the main radiation parameters in a wide frequency range, but they will also serve as a calibration of the UTR-2 effective area in the range of 25-33 MHz using astronomical methods. Currently the grating lobes of UTR-2 were taken into account only via calculations. Fig. 15 shows the average profiles of pulsars after two hours of integration at GURT subarray (observations were made during the years 2015-2016). Given that the average effective antenna area in the 30-70 MHz range is 2$\times$350 = 700 m$^2$ (after adding the signals of two polarizations), the obtained values of SNR (10-25) show an extremely high sensitivity of the antenna and the preamplification system of the GURT radio telescope, as well as the sensitivity of the ADR receiver.

\begin{figure}%[h]												% Fig Pulsars_GURT
\begin{center}
\includegraphics[width=1\linewidth]{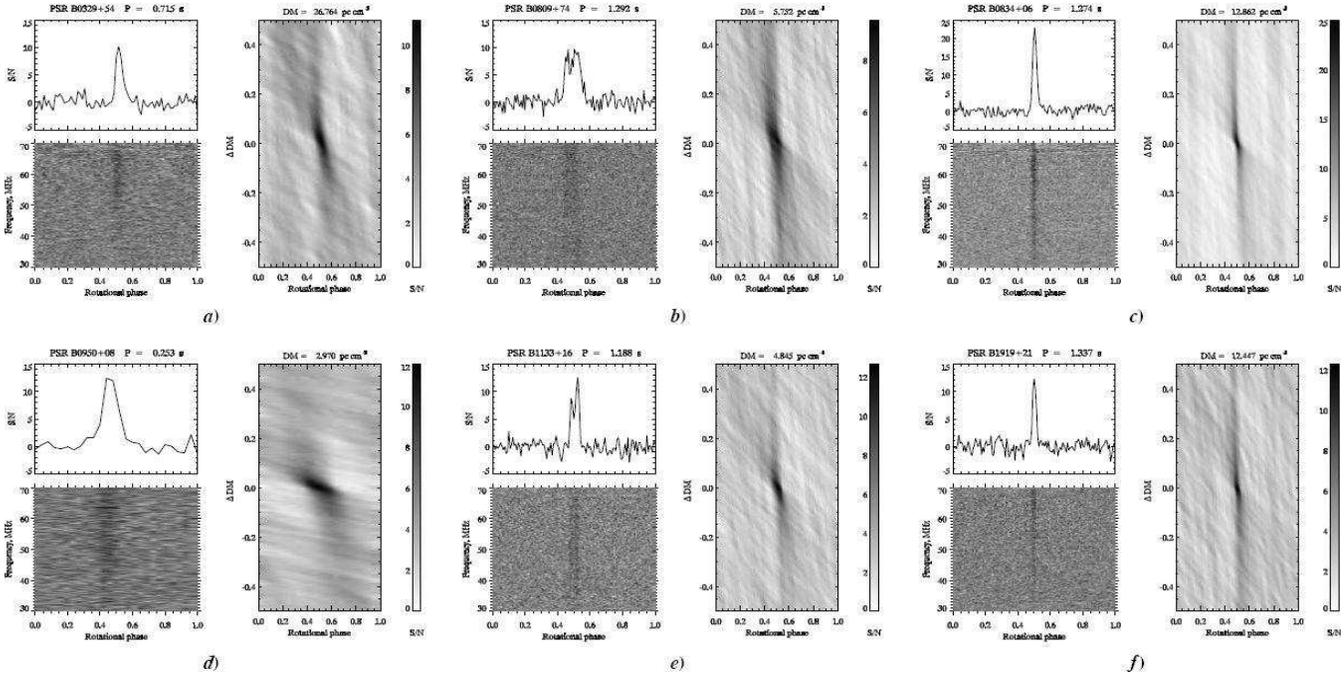} %100 percent
\end{center}
\caption{Spectrograms and the average profiles (left panel) and "DM vs Rotational phase" plane of some pulsars obtained at one GURT subarray. An integration time is 2 hours (ADR receiver).}
\label{Pulsars_GURT}
\end{figure}

% put the paragraph below in the conclusion section
Such a high sensitivity, broad frequency band and interference immunity open the way for solving a number of astrophysical problems both on individual subarrays, and with full radio telescope in coordination with other LF radio telescopes \cite{Konovalenko2016}.

\section{Conclusions and prospects}

Dedicated digital receivers for LF radio astronomy have shown a very high efficiency in astrophysical research. Low noise of the ADC requires the minimal gain of the radio telescope preamplification system. Hence the entire receiving signal path is highly linear, and the receiving system of the radio telescope has high RFI immunity.

A variety of available operation modes along with precise time and frequency referencing of the data in the receivers makes the experimental scheme flexible for solving particular astrophysical problems. Powerful RFI signatures typical of the LF range can be excised from the data due to high temporal and spectral resolutions of the receiver. Use of advanced RFI mitigation algorithms prevents the loss of valuable astrophysical information. 

The ADR receiver was equipped with a number of new features, including digital addition and subtraction of input signals, flexible choice of Fourier-sequence length, adjusting a signal level and a temporal delay between the channels on-the-fly. These features can be applied for adjusting the digital phasing system of GURT subarrays. This work can find the future application in the development of digital phase shifters for other LF radio telescopes.

Currently UTR-2 and URAN radio telescopes do not use all of their bandwidth reserve (up to 40 MHz and 50 MHz, respectively), provided by their dipole type and preamplification systems mainly due to frequency-dependent losses in their analog phasing systems. The larger the radio telescope dimensions, the stronger the total attenuation in the signal path at the upper operating frequency (70 dB for the UTR-2 and 35 dB for URAN-2). If the analog phasing system were replaced by the digital one, the frequency range would be expanded and the total effective area would be increase in the 30-50 MHz range. Additionally, formation of large number of antenna beams would become possible. These enhancements could enrich the variety of solvable astrophysical problems.

Numerous synchronous joint observations of Sun, pulsars, exoplanets and Saturn using the described digital receivers (DSPZ and ADR) at UTR-2, URAN, LOFAR and NDA are under way. Joint observations at LWA, NDA, URAN-2 and -3, UTR-2 and GURT are in progress as a part of a program of Jupiter radio emission study and ground-based support of Juno spacecraft mission.

\bibliographystyle{ws-jai}
\bibliography{JAI_Liter}

%\begin{thebibliography}{9}

%\bibitem[Blain {\it et al.}(2002)]{blain02} Blain, A. W., Smail, I., Ivison, R., Kneib, J.-P., Frayer, D. T. [2002] {\it Phys. Rep.} {\bf 369}, 111.

%...

%\end{thebibliography}
\end{document}